\documentclass[3p,times]{elsarticle}

\usepackage{amssymb}
\usepackage{graphicx,color}% Include figure files
\usepackage{epstopdf}
\usepackage[figuresright]{rotating}

\begin{document}

\begin{frontmatter}

%% Title, authors and addresses

%% use the tnoteref command within \title for footnotes;
%% use the tnotetext command for the associated footnote;
%% use the fnref command within \author or \address for footnotes;
%% use the fntext command for the associated footnote;
%% use the corref command within \author for corresponding author footnotes;
%% use the cortext command for the associated footnote;
%% use the ead command for the email address,
%% and the form \ead[url] for the home page:
%%
%% \title{Title\tnoteref{label1}}
%% \tnotetext[label1]{}
%% \author{Name\corref{cor1}\fnref{label2}}
%% \ead{email address}
%% \ead[url]{home page}
%% \fntext[label2]{}
%% \cortext[cor1]{}
%% \address{Address\fnref{label3}}
%% \fntext[label3]{}

%\dochead{}
%% Use \dochead if there is an article header, e.g. \dochead{Short communication}

\title{Modified DGLAP Evolution for Fragmentation Functions in Nuclei and QGP}

%% use optional labels to link authors explicitly to addresses:
%% \author[label1,label2]{<author name>}
%% \address[label1]{<address>}
%% \address[label2]{<address>}

\author[FIAS,Shanda]{Wei-Tian Deng}
\author[Shanda]{Ning-Bo Chang}
\author[LBNL]{Xin-Nian Wang}

\address[FIAS]{Frankfurt Institute for Advanced Studies (FIAS) Ruth-Moufang-Strasse 1, D-60438 Frankfurt am Main, Germany}
 \address[Shanda]{School of Physics, Shandong University, Jinan, Shandong 250100, China}
 \address[LBNL]{Nuclear Science Division, MS 70R0319, Lawrence Berkeley National Laboratory, Berkeley, California 94720}

\begin{abstract}
%% Text of abstract
Within the framework of generalized factorization of higher-twist contributions, including modification to splitting functions of both quark and gluon, we get and numerically resolve the medium-modified DGLAP (mDGLAP) evolution equations. With Woods-Saxon nuclear geometry and Hirano 3D ideal hydrodynamic simulations of hot medium, we study the medium modified fragmentation functions (mFF) in DIS and Au+Au collisions in RHIC. Our calculation imply that the parton density in hot medium produced in RHIC is about 30 times larger than cold nucleon.
\end{abstract}

\begin{keyword}
%% keywords here, in the form: keyword \sep keyword

%% MSC codes here, in the form: \MSC code \sep code
%% or \MSC[2008] code \sep code (2000 is the default)
Modified splitting functions; Modified Fragmentation Functions; nuclear modification factor
\end{keyword}

\end{frontmatter}

%%
%% Start line numbering here if you want
%%
% \linenumbers

%% main text
\section{Introduction}
\label{introduction}
Since jet quenching phenomenons are observed in RHIC, many phenomenological studies of it indicate a scenario of strong interaction between energetic partons and the hot medium with an extremely high initial parton density \cite{Wang:2003mm}.
The same phenomena are also predicted in deeply inelastic scattering (DIS) off large nuclei when the struck quark propagates through the target nuclei \cite{ww02}.

%Due to multiple scattering and induced gluon bremsstrahlung, the energetic parton propagating through medium will lose %a mount of energy in medium\cite{Gyulassy:1993hr,Baier:1996sk,Wiedemann:2000za,
%Gyulassy:2000er,Guo:2000nz,Wang:2001ifa} and therefor soften its final fragmentation functions.
%though the extracted parton density in cold nuclei is much smaller than that in the hot matter produced in the central %$Au+Au$ collisions at RHIC.

%Large transverse momentum hadrons in high-energy nucleon-nucleon collisions are produced
%through hard parton scattering with large transverse momentum transfer and the subsequent
%fragmentation of energetic partons into final hadrons due to soft gluon bremsstrahlung.
%In the leading logarithmic approximation, the resummation of these vacuum gluon bremsstrahlung will lead
%to a scale dependence of the jet fragmentation functions which can be described by
%the Dokshitzer-Gribov-Lipatov-Altarelli-Parisi (DGLAP) \cite{dglap} evolution equations in pQCD

In the presence of nuclear or hot QCD medium, the initially produced energetic partons will have to go
through multiple scattering and induced gluon bremsstrahlung before hadronization. The induced gluon bremsstrahlung effectively reduces the leading parton's
energy and softens the final hadron spectra or parton fragmentation functions. To take into account multiple induced gluon emissions, one can follow the resummation of gluon bremsstrahlung in vacuum and assume that multiple medium induced bremsstrahlung can be resummed in the same way to obtain the mDGLAP evolution equations for the modified fragmentation functions \cite{Guo:2000nz,Wang:2001ifa,Wang:2009qb},
\begin{eqnarray}
 \label{eq: modified DGLAP1}
 \frac{\partial \tilde{D}_q^h(z_h,\mu^2)}{\partial \ln \mu^2}&=&\frac{\alpha_s(\mu^2)}{2\pi}\int_{z_h}^1
 \frac{dz}{z}\left [ \tilde{\gamma}_{q\rightarrow qg}(z,\mu^2)\tilde{D}_q^h(\frac{z_h}{z},\mu^2)\right.
 +\left.\tilde{\gamma}_{q\rightarrow gq}(z,\mu^2)\tilde{D}_g^h(\frac{z_h}{z},\mu^2)\right ] ,\\
 \label{eq: modified DGLAP2}
  \frac{\partial \tilde{D}_g^h(z_h,\mu^2)}{\partial\ln \mu^2}&=&\frac{\alpha_s(\mu^2)}{2\pi}\int_{z_h}^1
 \frac{dz}{z}\left [ \sum_{q=1}^{2n_f}\tilde{\gamma}_{g\rightarrow q\bar q}(z,\mu^2)\tilde{D}_q^h(\frac{z_h}{z},\mu^2) \right.
 +\left. \tilde{\gamma}_{g\rightarrow gg}(z,\mu^2)\tilde{D}_g^h(\frac{z_h}{z},\mu^2)\right ] ,
\end{eqnarray}
where the modified splitting functions are given by the sum of the vacuum ones 
and the medium modification $\tilde \gamma _{a\rightarrow bc}(z,l_{T}^{2})=\gamma_{a\rightarrow bc}(z)+ \Delta \gamma_{a\rightarrow bc}(z,l_{T}^{2})$,

The first mDGALP evolution equation for quark in Eq.~(\ref{eq: modified DGLAP1}) is derived in Refs.~\cite{Guo:2000nz,Wang:2001ifa,Schafer:2007xh} in DIS off nuclei. The modification to the splitting functions for quark
\begin{eqnarray}
\label{eq:qg & gq}
 \Delta \gamma_{q\rightarrow qg}(z,x_{B},x_L,l_T^2)=\frac{1}{l_T^2}\left[\left( C_A \frac{1+z^2}{(1-z)_+} + C_F(1-z)(1+z^2) \right)T_{qg}^A(x_{B},x_L)+\delta(1-z)\Delta^{(q)} T_{qg}^A(x_{B},l_T^2)\right]
\times\frac{2\pi \alpha_s}{N_c f_q^A(x_{B})}\\
 %\Delta \gamma_{q\rightarrow gq}(z,x_{B},x_L,l_T^2)&=&\Delta \gamma_{q\rightarrow qg}(1-z,x_{B},x_L,l_T^2)
\end{eqnarray}
are obtained from the induced gluon bremsstrahlung spectra and therefore are related
to the twist-four nuclear quark-gluon correlation distribution $T^A_{qg}$ which essentially describe the amplitude of the second hard parton scattering and the induced gluon radiation.
The matrix element $\Delta^{(q)} T_{qg}^A$ in the second term in Eq.(\ref{eq:qg & gq}) comes from virtual corrections to the multiple parton scattering cross section. This term can be
constructed from the momentum sum rule (or momentum conservation) for mFF \cite{Guo:2000nz,Wang:2001ifa,Zhang:2003yn},
$\int dz z \Delta D_{q}^h(z,Q^{2})=0$.

In order to get the complete coupled mDGLAP evolution equations, we also consider multiple scattering and induced gluon bremsstrahlung for a gluon jet to get the mDGLAP evolution equation for gluon in Eq.~(\ref{eq: modified DGLAP2}). From gluon-gluon scattering matrix elements, one can obtain the medium modification to the splitting functions for gluon $\Delta \gamma_{g\rightarrow gg}$ and $\Delta \gamma_{g\rightarrow q \bar{q}}$ (see details in Ref.~\cite{Schafer:2007xh}).

In terms of the generalized jet transport parameter $\hat q$ in nuclear medium, we can express the quark-gluon correlation function $T^A_{qg}$ approximately as the integration over the parton's trajectory through the medium \cite{Osborne:2002st,CasalderreySolana:2007sw,Wang:2006qr}
\begin{equation}
\label{eq:Tqg and qhat}
\frac{2\pi\alpha_s}{N_c}\frac{T_{qg}^A(x_B,x_L)}{f_q^A(x_B)} \approx
 \int dy^- \hat q_{F}(y) 4 \sin^{2}(x_{L}p^{+}y^{-}/2) \, .
\end{equation}

Given the initial conditions of fragmentation functions at initial energy scale $\tilde{D}_a(Q_0^2)$, We can numerically solve the coupled mDGLAP evolution equations using modified HOPPET \cite{hoppet} Fortran 95 package in LO.
Such initial conditions in principle should be different in medium and vacuum. To take into account
medium modification to the fragmentation functions at the initial scale $Q_{0}^{2}$, we will assume in
this study \cite{Wang:2009qb}$ \tilde D_{a}(Q_0^2)=D_{a}(Q_0^2)+\Delta D_{a}(Q_{0}^{2})$,
where $D_{a}(Q_0^2)$ is the vacuum fragmentation function and $\Delta D_{a}(Q_{0}^{2})$ is
generated purely from medium via the mDGLAP starting at $\mu^{2}=0$.

\section{Modified Fragmentation Function in DIS}
\label{section:DIS}
To calculate the mFF in semi-inclusive DIS off a nucleus, we employ the Woods-Saxon nuclear geometry. Considering the initial quark jet produced at $y_{0}$ that travels along a direction with impact parameter $b$, we assume that the jet transport parameter along the quark jet trajectory is proportional to the nuclear density $\hat q(y,b)=\hat q_{0} \rho_{A}(y,b)/ \rho_{A}(0,0)$.

If we neglect the nuclear and impact parameter dependence of the nuclear quark distribution
function, the photon-nucleon cross section that produces a quark at $(y_{0},b)$ is proportional
to the nuclear density distribution $\rho_{A}(y_{0},b)$. Then the averaged mFF should be
\begin{equation}
 \tilde{D}(z)=\langle\tilde{D}(z,y_0,b)\rangle=\frac{\pi}{A}\int_{0}^{\infty} db^{2} \int_{-\infty}^{\infty} dy_0 \tilde{D}(z,y_0,b) \rho_A(y_0,b).
\end{equation}
In order to calculate the $\tilde{D}(z,y_0,b)$ for a quark produced at location $(y_0,b)$, the path integral in the modified splitting functions should be replaced by the following
\begin{equation}
\int 4dy^- \hat q(y)\sin^{2}(x_{L}p^{+}y^{-}/2) =
 \frac{\hat q_{0}}{\rho_{A}(0,0)} \int_{y0}^{\infty} 4dy^- \rho_A(y,b)
\sin^2 \left[ \frac{l^2_{T} (y-y_0)}{4q^- z(1-z)} \right]
\end{equation}

\begin{figure}
  \centering
 \includegraphics[width=0.7\textwidth]{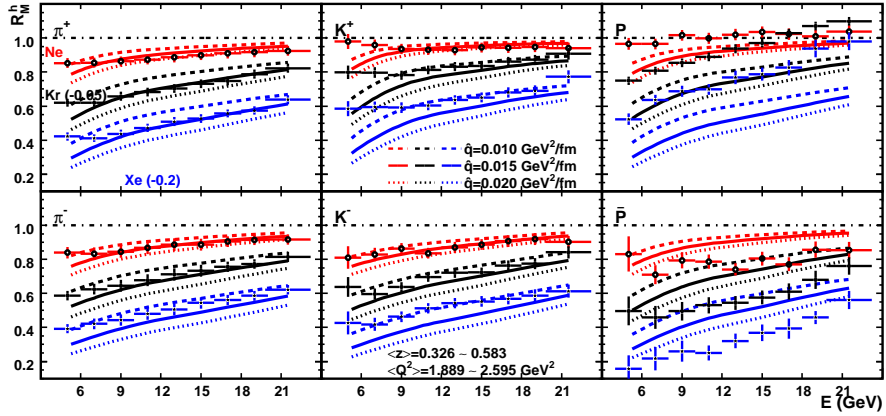}
   \caption{The energy dependency of the nuclear modification factors with different values of the jet transport
  parameter $\hat q_{0}$ compared with the HERMES \cite{HERMES}
  data for $Ne$, $Kr$ and $Xe$ targets. For clear presentation the modification factors for different targets have
  been shifted vertically by some value($Kr$ by -0.05 and $Xe$ by -0.2).}
  \label{fig:HERMES_nu}
\end{figure}

Shown in Fig.~(\ref{fig:HERMES_nu}) are the calculated nuclear modification factors of $\pi^{\pm}$, $K^{\pm}$
and $p(\bar p)$ with different values of jet transport parameter $\hat q_{0}=0.015\pm 0.005$
GeV$^{2}$ as compared to the HERMES experimental data \cite{HERMES}:
\begin{eqnarray}
R_M^h(z,\nu)=\left(\frac{N^h(z,\nu)}{N^e(\nu)}|_A\right) / \left(\frac{N^h(z,\nu)}{N^e(\nu)}|_D\right)
=\left(\frac{\Sigma e_f^2q_f(x)D_f^h(z)}{\Sigma e_f^2q_f(x)}|_A\right) / \left(\frac{\Sigma e_f^2q_f(x)D_f^h(z)}{\Sigma e_f^2q_f(x)}|_D\right).
\end{eqnarray}

 Illustrated in Fig.~(\ref{fig:HERMES_nu}), the medium modification of the mFF gradually disappears as the initial jet energy $E$ increases. The agreement between our theory calculations and experimental data is generally good except at lower energy where hadronic absorption might become important. We can see significant different modification at HERMES for $p$ and $\bar{p}$, this may be because of the process of quark-antiquark annihilation in twist-4 double scattering and the asymmetry of $q$ and $\bar{q}$ in nuclei \cite{Schafer:2007xh,Zhang:2007zzr}, which we have not put into mDGLAP yet.

As we have discussed, the initial condition for mFF in the medium at $Q_{0}^{2}=1$ GeV$^{2}$ is different from which in vacuum. Therefore, most of the medium modification to the mFF come from mDGLAP evolution at low $Q^{2}$ while contribution from high $Q^{2}$ region is power-suppressed. This will lead
to a very weak $Q^{2}$ dependence as shown in the left panel of Fig.~\ref{fig:HERMES_q}. The calculated suppression factors are almost independent of $Q^{2}$, consistent with the experimental data \cite{HERMES}. If one has chosen
the initial condition at $Q_{0}^{2}$ as the same as the vacuum one, one would obtain a too strong $Q^{2}$ dependence of modification factor.
%\begin{figure}
%  \centering
% \includegraphics[width=0.8\textwidth]{z_HERMES_all.eps}
%  \caption{The modified multiplicity ratios as a function of $z$ with different values of the jet transport parameter %$\hat q_{0}$ compared with the HERMES \cite{HERMES} data for $Ne$,$Kr$ and $Xe$ targets. For clear presentation the %modification factors for different targets have been shifted vertically by
%  some value($Kr$ by -0.05 and $Xe$ by -0.2).}
%  \label{fig:HERMES_z}
%\end{figure}

\section{Modified Fragmentation Function in Au+Au Collisions}

One can extend the calculation for medium modified parton fragmentation functions in DIS to hot medium \cite{CasalderreySolana:2007sw,Majumder:2007ae} like QGP or hot hadronic matter created in high-energy heavy-ion collisions. To take into account both the the longitudinal and transverse expansion of the hot matter we use a 3D ideal hydrodynamic simulations \cite{Hirano:2005xf,Hirano:2007ei} which give us information of the temprature, energy density, the fraction of the hadron phase and so on, on each step of the hot matter evolution.

The jet transport parameter in hot medium at given time $\tau$ and local position $r$ is assumed to include the contribution from both the  QGP phase and hadronic phase \cite{Chen:2010te}
\begin{equation}
\label{Eq:q-hat-qgph}
\hat{q} (\tau,\textbf{r})= \hat{q}_0\frac{\rho^{QGP}(\tau,\textbf{r})}{\rho^{QGP}(\tau_{0},0)}  (1-f) + \hat q_{N} \frac{\rho_{h}(\tau,\textbf{r})}{\rho_{N}} f \,,
\end{equation}
where f is the fraction of the hadronic phase, $\hat{q}_0$ is the jet transport parameter at the center of the bulk medium in the QGP phase at its formation time $\tau_{0}$.  $\rho_{h}(\tau,\textbf{r})$ is the number density of the hadron resonance gas, $\rho_{N}=n_{0}\approx 0.17$ fm$^{-3}$ is the nucleon density in the center of a large nucleus and $\hat q_{N}=0.015$ GeV$^2/$fm is the jet transport parameter in cold nuclear matter we got in last section.

Then we can calculate the mFF $\tilde{D}_{p}^{h}(z_h,Q^2,E,\textbf{r},\phi,\textbf{b})$ in hot medium, and average it over the initial parton production position and the out going directions. Assume the parton production cross section is proportional to the overlap function, we can get
\begin{eqnarray}
\langle \tilde D_{p}^{h}(z_h,Q^2,E,\textbf{b})\rangle =
\frac{\int d\phi d^2{\textbf{r}} t_{A}(|\textbf{r}+\textbf{b}|)t_{A}(|\textbf{r}-\textbf{b}|)\tilde{D}_{p}^{h}(z_h,Q^2,E,\textbf{r},\phi,\textbf{b})} {2\pi \int d^2{\textbf{r}}t_{A}(|\textbf{r}+\textbf{b}|)t_{A}(|\textbf{r}-\textbf{b}|)}.
\label{eq:frag}
\end{eqnarray}

\begin{figure}
  \centering
 \includegraphics[width=0.35\textwidth]{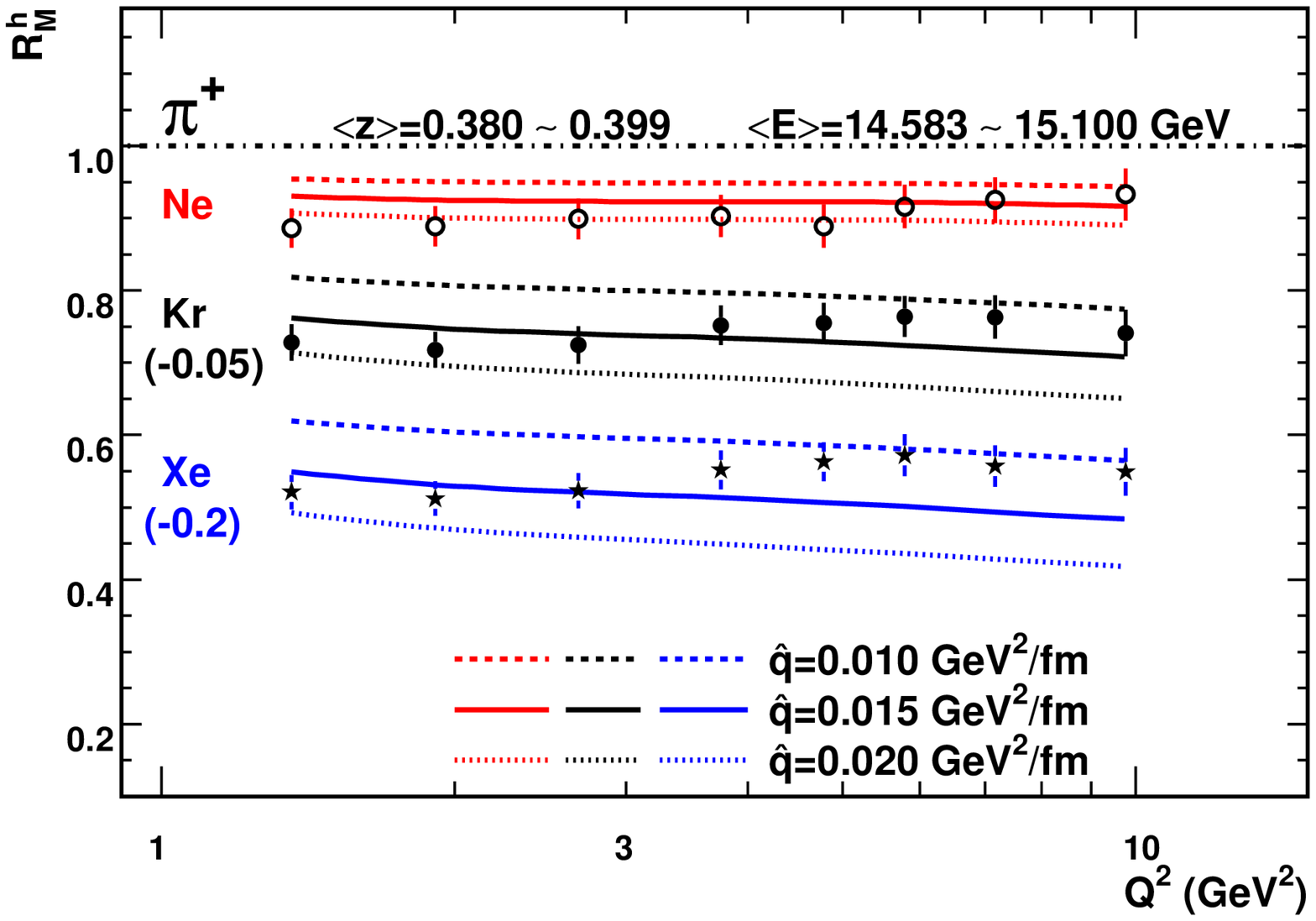}
 \includegraphics[width=0.35\textwidth]{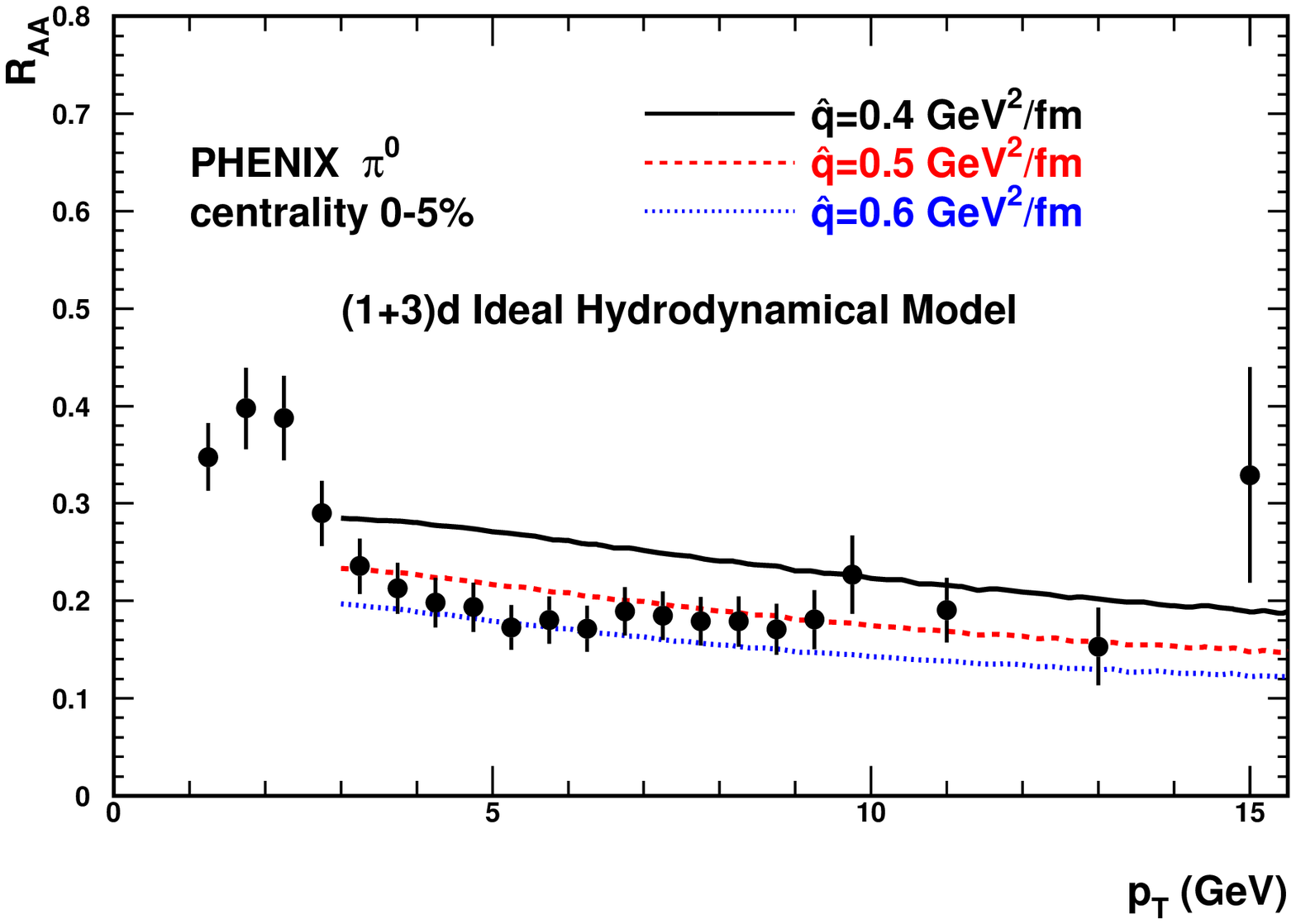}
  \caption{left: Comparison of the modified multiplicity ratios as a function of $Q^{2}$ at fixed value of $z$ and jet energy $E$ with the HERMES \cite{HERMES} data for $Ne$, $Kr$ and $Xe$ targets. For clear presentation the modification factors for different targets have been shifted vertically by some value($Kr$ by -0.05 and $Xe$ by -0.2).
right: comparison of the nucleon modification factor $R_{AA}$ for $\pi^0$ in $0-5\%$ biased events.}
  \label{fig:HERMES_q}
\end{figure}

If we neglect nuclear effect such as the shadowing effect on the initial parton distribution function, we can get the nuclear modification factor $R_{AA}$ for a fixed impact parameter $\textbf{b}$
\begin{eqnarray}
R_{AA}(\textbf{b})=\frac{d\sigma_{AB}^h/dyd^2p_Td^2\textbf{b}}{T_{AA}(\textbf{b})
d\sigma_{pp}^h/dyd^2p_T}
=\frac{f^p(x_1,x_2)\otimes \mathrm{d}\sigma \otimes \langle \tilde D_{p}^{h}(z_h,Q^2,E,\textbf{b})\rangle} {f^p(x_1,x_2)\otimes \mathrm{d}\sigma \otimes D(z_h,Q^2)}.
 \label{eq:rab}
\end{eqnarray}
Shown in the right panel of Fig.~\ref{fig:HERMES_q} is the comparison of our result about the medium modification factor for top $5\%$ centrality Au+Au collisions to the PHENIX data \cite{Adare:2008qa}. The jet transport parameter at the center of the bulk medium in the QGP phase at its formation time $\hat{q}_0=0.5 \pm 0.1$ GeV$^2/$fm can fit the experiment data in error bar.

%\begin{figure}
%  \centering
% \includegraphics[width=0.6\textwidth]{Raa_Hir.eps}
%  \caption{}
%  \label{fig:Raa_Hirano}
%\end{figure}

\section{Conclusions}
We have got the coupled mDGLAP evolution equations within the framework of generalized factorization of higher-twist contributions. By solving the mDGLAP equations numerically, we study the nuclear modified fragmentation functions both in cold nucleon and hot medium produced in RHIC. From our results, one is suggested that the gluon density in QGP is about 30 times larger than which in cold nucleon.

\section*{Acknowledgements}
We would like to thank Xiao-Fang Chen and Zhe Xu for their helpful discuss. This work is supported by the Director, Office of Energy Research, Office of High Energy and Nuclear Physics, Divisions of Nuclear Physics, of the U.S. Department of Energy under Contract No. DE-AC02-05CH11231. W.-T. Deng was financialy supported by Helmholtz International Center for FAIR within the framework of the LOEWE program launched by the State of Hesse. N.-B. Chang was financialy supported by National Natural Science Foundation of China under Project Nos. 10975092.

%% The Appendices part is started with the command \appendix;
%% appendix sections are then done as normal sections
%% \appendix

%% \section{}
%% \label{}

%% References
%%
%% Following citation commands can be used in the body text:
%% Usage of \cite is as follows:
%%   \cite{key}         ==>>  [#]
%%   \cite[chap. 2]{key} ==>> [#, chap. 2]
%%

%% References with BibTeX database:

\bibliographystyle{elsarticle-num}
%\bibliography{<your-bib-database>}

%% Authors are advised to use a BibTeX database file for their reference list.
%% The provided style file elsarticle-num.bst formats references in the required Procedia style

%% For references without a BibTeX database:

% \begin{thebibliography}{00}

%% \bibitem must have the following form:
%%   \bibitem{key}...
%%

% \bibitem{}

% \end{thebibliography}

\end{document}